# Inclusion of Machine Learning Kernel Ridge Regression Potential Energy Surfaces in On-the-Fly Nonadiabatic Molecular Dynamics Simulation

*Deping Hu,[†,‡] Yu Xie,[†] Xusong Li,[†,‡] Lingyue Li,[‡] and Zhenggang Lan\*[†,‡]*

[†]CAS Key Laboratory of Biobased Materials, Qingdao Institute of Bioenergy and Bioprocess Technology, Chinese Academy of Sciences, Qingdao 266101, China

[‡]University of Chinese Academy of Sciences, Beijing 100049, China

AUTHOR INFORMATION

**Corresponding Author**

*Email: lanzg@qibebt.ac.cn

**ABSTRACT**: We discuss a theoretical approach that employs machine learning potential energy surfaces (ML-PESs) in the nonadiabatic dynamics simulation of polyatomic systems by taking 6-aminopyrimidine as a typical example. The Zhu-Nakamura theory is employed in the surface hopping dynamics, which does not require the calculation of the nonadiabatic coupling vectors. The kernel ridge regression is used in the construction of the adiabatic PESs. In the nonadiabatic dynamics simulation, we use ML-PESs for most geometries and switch back to the electronic structure calculations for a few geometries either near the $S_1/S_0$ conical intersections or in the out-of-confidence regions. The dynamics results based on ML-PESs are consistent with those based on CASSCF PESs. The ML-PESs are further used to achieve the highly efficient massive dynamics simulations with a large number of trajectories. This work displays the powerful role of ML methods in the nonadiabatic dynamics simulation of polyatomic systems.

Nonadiabatic transition plays an important role[1-2] in plenty of photochemical and photophysical processes, such as the photostability of biological polyatomic systems,[3-5] the photoisomerization of organic conjugated alkenes and their derivatives,[6-7] and the photoreactions of transition metal complexes.[8-9] In the vicinity of conical intersections (CIs), the strong interstate coupling drives ultrafast nonadiabatic processes, in which the Born-Oppenheimer approximation breaks down.[1-2] Various nonadiabatic dynamics methods have been developed, including quantum dynamics,[1-2,10-12] and mixed-quantum-classical or semiclassical dynamics.[13-21]

The trajectory surface hopping (TSH) method[22-25] is a widely used mixed quantum-classical method for the treatment of nonadiabatic dynamics. Different methods have been proposed to define the hopping probability in the TSH dynamics. One approach assumes that the hopping probability is relevant to the electronic evolution governed by time-dependent quantum mechanics. One widely used approach that belongs to this theoretical framework is the fewest-switches algorithm proposed by Tully.[22] Alternatively, it is also possible to use a predefined probability, for instance, based on the Landau-Zener formulism,[26] in the TSH calculations. Along this line, Zhu and Nakamura developed the Zhu-Nakamura method to calculate the nonadiabatic transition probability.[27-28] Later, Zhu and coworkers proposed the simplified Zhu-Nakamura formulism to calculate the hopping probability, which does not require the calculation of nonadiabatic coupling vectors (NACVs) in the TSH simulation.[29-31] Although it is a rather simplified method, this approach shows promising results with respect to those obtained from Tully's surface-hopping dynamics method.[31]

Independent of the dynamical method chosen to treat nonadiabatic processes, one key step is to obtain the potential energy surfaces (PESs) and their couplings. It is possible to build the PESs before the dynamics run.[32] This so-called "PES-fitting" method generally assumes the particular functions of the PESs and then fits the relevant parameters on the basis of experimental data or electronic structure calculations. A large amount of electronic structure calculations over an extensive high-dimensional grid should be performed to build the database for fitting, which becomes a difficult task with an increasing system size. It is also not trivial to define the suitable active coordinates in the PES construction for large systems.

Alternatively, when mixed-quantum-classical dynamics is considered, it is possible to perform the electronic structure calculations in an on-the-fly manner. During the past decade, great efforts were made to combine the direct dynamics with mixed quantum-classical dynamics for the atomic treatment of the nonadiabatic dynamics of complex systems with all degrees of freedom.[32-41] On-the-fly TSH calculations may suffer from an enormous computational cost because the electronic-structure calculations are performed at each time step of nuclear motion. This computational cost prevents the simulation of a large number of trajectories, particularly for complex systems.

Recently, the machine learning (ML) approach has been widely used in various fields,[42] such as brain-computer interfaces, pattern recognition, recommender systems and robotics. The supervised ML methods have been applied very successfully in many fields of chemistry,[43-44] including the structure-property relation,[45-48] the construction of PESs,[49-63] and the excitonic dynamics.[64] In the study of ground-state dynamics, various ML approaches have been proposed to construct the PESs, including kernel ridge regression (KRR),[65-67] artificial neural network (ANN),[43,49-51,54,56-58,61-63,68-75] and Gaussian process models.[76] Although ML-PESs have started to become important in ground-state dynamics simulations, the employment of ML-PESs in the excited-state nonadiabatic dynamics simulation is still limited.[61,64,73,77]

In recent years, some efforts have been made to construct the diabatic PESs and the interstate couplings of middle-sized molecules.[61,64,73,77-81] These efforts either assume analytical PES functions in the fitting procedure or employ ML approaches to build the PES functions. This approach provides the possibility to simulate the full-dimensional nonadiabatic dynamics of complex systems with accuracy and efficiency. Most of these studies have tried to fit the diabatic PESs and couplings because the physical properties become smooth in a diabatic representation. However, the construction of the diabatic model is extremely challenging in complex systems. In addition, it is not easy to define suitable coordinates in the fitting procedure.

In this work, we attempted to combine the ML-PESs and the nonadiabatic dynamics simulation. We utilized a rather simple approach to perform the TSH simulation on the basis of the Zhu-Nakamura algorithm.[29-31] As mentioned in a previous study,[31] the Zhu-Nakamura method can provide a reasonable description of the nonadiabatic dynamics of polyatomic systems. The further approximation of the Zhu-Nakamura algorithm in the TSH method does not need NACVs.[29-31] We chose 6-aminopyrimidine (6AP)[82] as a typical system to explain our approach. After the construction of the molecular descriptor via the Coulomb matrix (CM),[46-47,64] we attempted to employ the KRR[65-67] approach to build the adiabatic PESs. After we carefully controlled the sampling and fitting procedure, the fitted PESs performed very well along the pathway. In the nonadiabatic dynamics simulation, we mainly used the resulting ML-PESs for most structures. To avoid possible ML-PES fitting errors relevant to the double-cone topology of the PESs, we switched back to the electronic structure calculations for only a few geometries when the trajectory enters areas in the vicinity of the CI seam. This approach provides an excellent balance between computational cost and accuracy. As a complementary study, we also ran two sets of on-the-fly TSH simulations based on both Zhu-Nakamura's and Tully's methods to check the accuracy of the former approach.

As a pyrimidine heterocycle and a reduced model of adenine, 6AP (Figure 1) was chosen as the prototype because its nonadiabatic dynamics was extensively studied in previous investigation.[82]

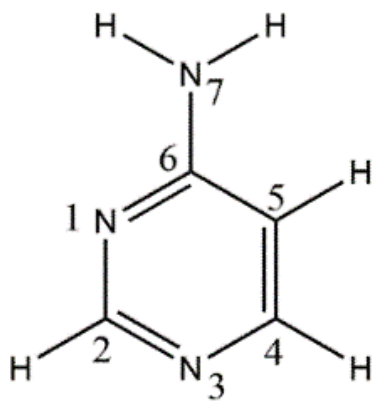

**Figure 1.** Molecular structure of 6AP and atomic labels.

An adaptive learning strategy was used in the construction of the PESs and dynamics simulation. Next, a step-by-step protocol is provided, and more technical details are given in Supporting Information.

(1) 100 initial conditions near the $S_0$ ground state minimum (GSM) were generated using Wigner sampling at both 0 and 1000 K. The high-temperature (1000 K) sample was used to broaden the sampling space. On-the-fly Born-Oppenheimer molecular dynamics (BOMD) simulations initialized from $S_1$ were performed based on these initial conditions. The trajectory ran with 0.5 fs time step up to 2000 steps at 0 K (and 100 steps at 1000 K), while some trajectories may crash during the dynamics run. The snapshots in the dynamics run were collected to form our initial data set library (196754 6AP conformations). A part of this data set was used to build the trial training data set (95775 6AP conformations). Then, the hierarchical agglomerative clustering algorithm[42] was used to divide these data into 700 groups. If a cluster contained fewer than 100 geometries, then we selected all of the geometries. Otherwise, we selected at most 100 geometries from each group. All selected geometries formed the first training data set (60243 6AP conformations). We obtained preliminary PESs of 6AP by the KRR.

(2) Another 50 initial conditions were generated with Wigner sampling at 0 K. The Zhu-Nakamura nonadiabatic dynamics simulation starting from the $S_1$ state was performed based on the preliminary ML-PESs obtained at Step 1. After the simulation, we obtained the excited-state lifetime and hopping geometries. The CI optimization starting from representative hopping geometries gave us the minimum-energy $S_0/S_1$ CIs. We constructed the linear-interpolated reaction pathway from $S_0$ to the CIs based on the ML-PESs and compared them with those obtained using CASSCF method. We found that the ML-PES results are consistent with the CASSCF results, except for the $S_1$ barrier region caused by the $S_2/S_1$ crossing. Next, we sampled 30 initial geometries near this region and performed a very short-time dynamics simulation (BOMD: 150 steps) initiated from $S_2$, in which the initial velocities were set to 0. We added geometries generated from the short-time dynamics simulation (4254 6AP conformations) into our original data set library. At last, the refined data set library had a total of 201008 6AP conformations.

(3) A vast number of geometries were selected randomly to define the preliminary training data set. We performed the clustering analysis over these data. Next, we randomly selected the same number of geometries from each cluster to define the final training data set (65316 6AP conformations) and retrained the PESs again with KRR. The clustering trick largely reduced the computational cost in the training step without losing accuracy.

(4) Because the fitting involves the determination of two parameters, $\sigma$ and $\lambda$, a validation data set (39971 6AP conformations selected from the data set library) was used to determine them. At the end, the obtained new ML-PESs were used to test their predication ability on the basis of the test data set (61008 6AP conformations selected from the data set library). Certainly, the training, validation, and test data sets should not have any overlap of the data points.

(5) To check the accuracy of the newly trained ML-PESs, another 100 initial conditions were generated, and the Zhu-Nakamura nonadiabatic dynamics simulation was performed based on both the newly trained ML-PESs and the pure on-the-fly CASSCF PESs. Because the prediction by KRR is an interpolation method, the PESs of a particular geometry can only be correctly predicted if this geometry lies in the sampling area. However, in the dynamics process, some trajectories may go out of the sampling region, which could not be correctly predicted with the KRR method due to its limitation to interpolation prediction. Therefore, for these geometries out of the sampling areas, CASSCF calculations were invoked. In our simulation, such situation is determined by the so-called out-of-confidence geometry that was determined if its CM was not located in the 95% confidence interval calculated with the CMs in the training set. A slightly different idea was suggested in previous work.[53] Because the PESs display a double-cone topology in the vicinity of the CI region, the PESs near the CI seam are not smooth functions of the nuclear geometry. The current fitting PESs cannot give a fully satisfied description on such a cusp. To avoid this problem, we switched back to the CASSCF calculation when the energy difference between the different states was <0.1 eV in the dynamics simulations. This approach should help us obtain more reliable results near the CI. The total number of the geometries using the CASSCF results instead of the ML-PESs accounts for 3.9% of all of the geometries in the dynamics simulations.

(6) After we made sure that the ML-PESs worked well, another dynamics simulation with 1000 trajectories was performed based on the ML-PESs. Still, for a few geometries that are out of the reliable confidence interval or are very close to the $S_0/S_1$ CIs, we switched to the CASSCF calculations.

Before the discussion of the dynamical simulation results based on ML-PESs, we first checked the accuracy of the ML-PESs. Figure 2 shows the distributions of the test errors in the learning procedure for the ML-PESs of 6AP. The mean absolute errors (MAEs) are also presented.

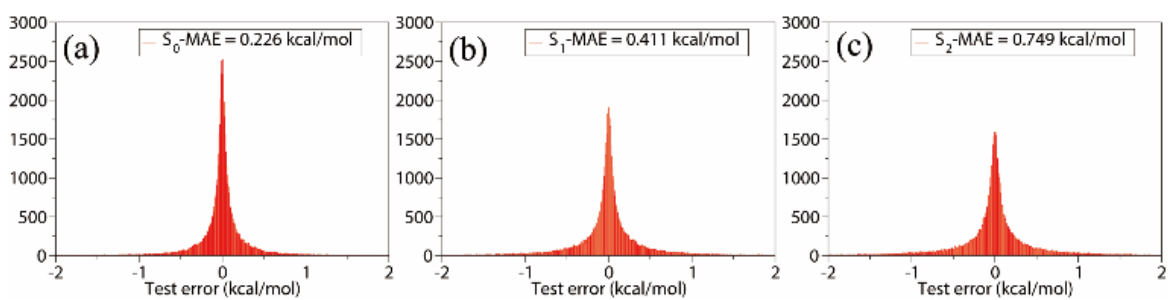

**Figure 2.** Distributions of the test errors for the adiabatic energy of state (a) $S_0$, (b) $S_1$, and (c) $S_2$ based on ML-PESs.

We can see that the ML-PESs accurately predict the energies of all the adiabatic electronic states ($S_0$, $S_1$, and $S_2$). Notice that the test errors increase with an increase in the adiabatic electronic state order (with MAE from 0.226 to 0.749 kcal/mol). A possible explanation is because the density of states is very high for the excited states, and even at the Franck-Condon region, different states may strongly mix with each other and the PES crossings are easily formed in the high-dimensional space. These effects result in more nonsmooth areas on the excited-state PES, which further leads to fitting errors. When the state order increases, such a mixture between different characters becomes stronger, and thus the fitting errors become larger. However, the MAEs are still low enough (< 1 kcal/mol) to ensure the possibility of using the ML-PESs on all states involved in the dynamics simulations.

Two CIs, $C_2$-CI and $C_4$-CI, are responsible for the nonadiabatic decay of 6AP, whose geometries are distinguished by the ring deformation at the $C_2$ and $C_4$ atom, respectively. This result is consistent with previous work.[82] As shown in Figure 3, the potential energy curves of the reaction pathways from the GSM to the two CIs are well predicted using the ML-PESs compared with the CASSCF PESs.

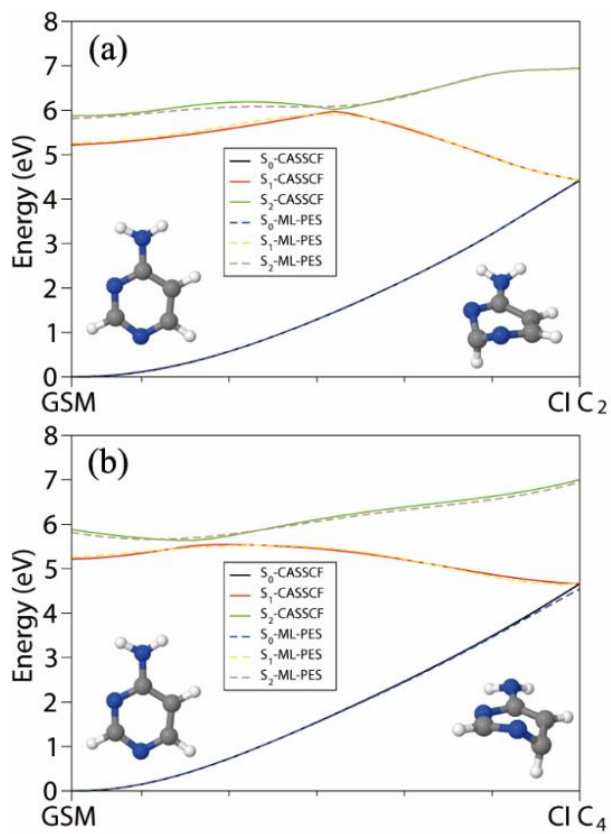

**Figure 3.** Linearly interpolated reaction pathways from the GSM to (a) $C_2$ and (b) $C_4$ CIs based on the CASSCF calculation (solid line) and ML-PESs (dashed line), respectively. All of the GSM and CIs are optimized at the CASSCF level.

For all states, the potential energy curves based on the ML-PESs and CASSCF are very similar along the reaction pathways, even in the vicinity of the regions where two states come closer. The good performance of the ML-PESs may be attributed by the fact that the BOMD was employed in sampling, in which the system accesses the $S_0/S_1$ CI region several times. However, the double-cone topology of PESs in the branching space may not always be reproduced by the KRR along the whole CI seam. Thus we switched back to electronic structure calculations when the trajectory accesses the CI region in the nonadiabatic dynamics. Note that the additional geometry sampling near the barrier on the $S_1$ state or $S_1/S_2$ crossing is also necessary to generate the reliable ML-PESs for reaction pathway calculations and nonadiabatic dynamics simulation.

When the ML-PESs are employed to treat most geometries in the Zhu-Nakamura TSH nonadiabatic dynamics simulation, the time-dependent average fractional occupations of the adiabatic electronic states are shown in Figure 4a. The $S_0$ time-dependent fractional occupation is fitted by an exponential function, $f(t) = 1 - \exp(-(t - t_a)/t_b)$. The excited-state lifetime is given by $\tau = t_a + t_b$, where $t_a$ = 64 fs, $t_b$ = 271 fs, and $\tau$ = 335 fs, consistent with previous theoretical study with an excited-state lifetime of 416 ± 150 fs.[82]

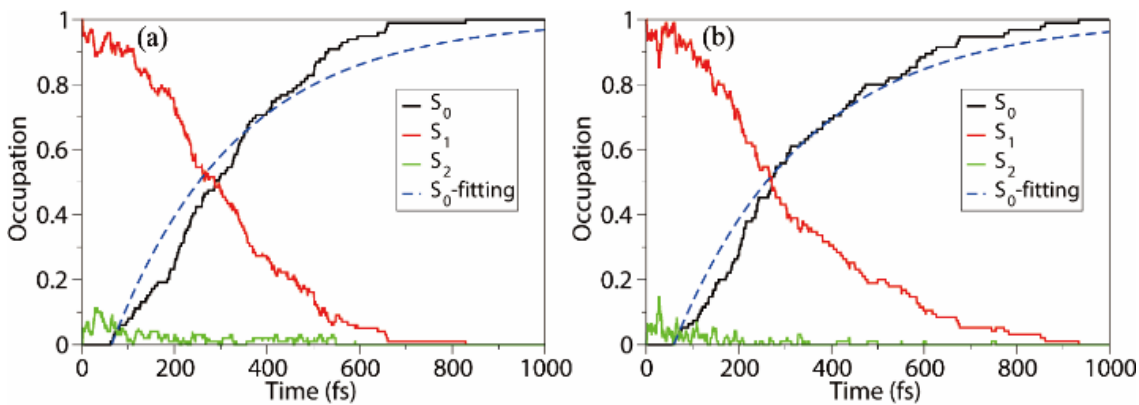

**Figure 4.** Time-dependent average fractional occupations of the adiabatic electronic states for the Zhu-Nakamura nonadiabatic dynamics with 100 trajectories initiated from $S_1$ using (a) ML-PESs and (b) CASSCF PESs. The fitting results of the $S_0$ occupation are also given with dashed lines.

The benchmark result based on the on-the-fly dynamics calculations at the CASSCF level is shown in Figure 4b. The time-dependent state occupations agree well with those obtained based on the ML-PESs, even for a small population change in the $S_2$ state. The excited-state lifetime is also estimated giving $t_a = 60$ fs, $t_b = 287$ and $\tau = 347$ fs. Thus the population dynamics show almost the same excited-state lifetime independent of whether we use the ML-PES or CASSCF PES.

The nonadiabatic dynamics simulations based on our current approach and pure CASSCF PESs provide a similar geometrical distribution at hopping events (Figure 5). Although the former distribution is slightly broader, it still reproduces the major geometrical features at hopping events in the dynamics process. Clearly, the major reaction channel is via the $C_4$-CI, while the secondary channel is governed by the $C_2$-CI or the $N_3$ out-of-plane motion. This result is consistent with previous work.[82] (Please note that the optimization of the hopping geometries with the $N_3$ out-of-plane motion will stop at either $C_4$-CI or $C_2$-CI, which is consistent with previous work.)

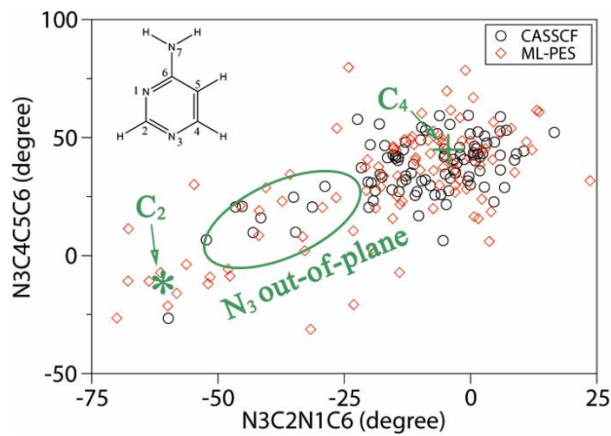

**Figure 5.** Geometry distributions of the key internal coordinates (dihedral N3C2N1C6 and N3C4C5C6) at hops in the dynamics simulation based on the CASSCF PESs (circle) and ML-PESs (diamond), respectively. The positions of the two CIs ($C_2$ and $C_4$) are also given.

Because of the good performance of the ML-PESs in the dynamics simulations, we believe that the ML-PESs in this study are reliable enough for even massive dynamics simulations. Thus the dynamics simulations with another 1000 trajectories were performed, as shown in Figure 6. Compared with the dynamics results with 100 trajectories, a clearer overview of the dynamics feature can be observed, such as a smoother change in the state occupation (Figure 6a) and a clearer distribution of the hopping geometries (Figure 6b).

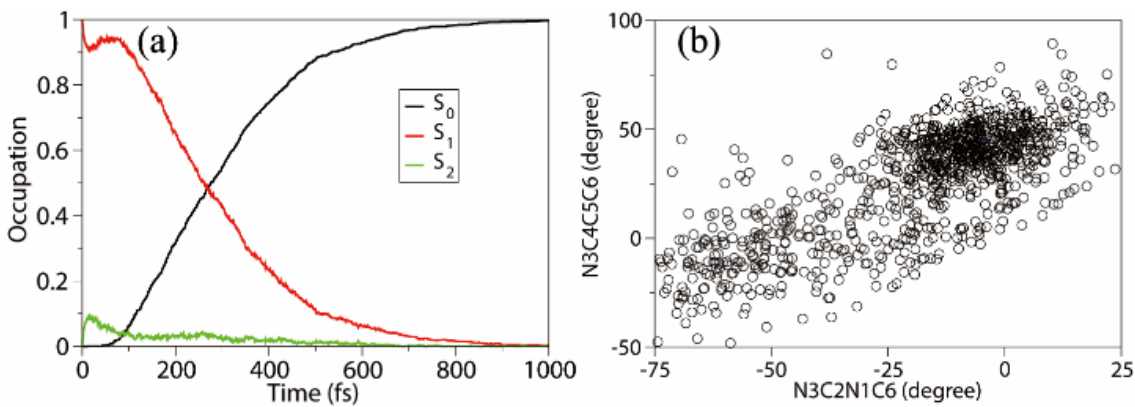

**Figure 6.** Nonadiabatic dynamics results with 1000 trajectories initiated from $S_1$ using the ML-PESs: (a) Time-dependent average fractional occupations of the adiabatic electronic states and (b) the geometry distributions of dihedral N3C2N1C6 and N3C4C5C6 at hops in the simulation process.

In many published studies of the TSH dynamics on model PESs,[22,29,83] a large number of trajectories (generally more than 1000 even 5000) was normally required to ensure fully convergent calculations. This trajectory number implies that the on-the-fly TSH dynamics with 100-200 trajectories basically gives a rather approximated description of the nonadiabatic dynamics of polyatomic systems. If we wish to provide a more reliable description on the final branching ratio comparable to experimental data, then thousands of trajectories or more should be computed.[84] To improve the statistical accuracy of the TSH dynamics, it is very important to develop reliable approaches to go beyond the current limits of on-the-fly simulations and to compute many more trajectories. The current work provides some useful ideas along this line.

The reliable PES fitting is much more challenging in excited-state dynamics than in ground-state dynamics. Most ground-state molecular dynamics essentially describe the thermal equilibrium distribution. The trajectories often stay in the deep well and exceed the barrier much less, resulting in the good clustering feature in phase space. However, the ultrafast excited-state dynamics is in a highly nonequilibrium state and may also involve strongly distorted geometries, such as a large ring deformation in our current system. The construction of excited-state PESs, which reliably describes the nontrivial geometrical deformation, the low barrier on the flat PES, and the highly nonequilibrium ultrafast dynamics, is not trivial. The current work provides a useful approach to build multidimensional PESs to treat the more general nonadiabatic dynamics of polyatomic systems.

During the last few decades, considerable theoretical efforts were made to develop other rigorous semiclassical dynamics, such as quantum-classical Liouville equations[16] and semiclassical initial value representation,[15] for a more accurate description of the nonadiabatic dynamics, while these methods normally require a tremendous number of trajectories. Thus the current effort to construct multidimensional PESs opens a new possibility to treat the nonadiabatic dynamics of polyatomic systems with rigorous semiclassical dynamics.

The current work is mainly based on Zhu-Nakamura TSH dynamics. It is more challenging to employ supervised ML approaches to perform Tully's surface hopping dynamics that requires an estimation of the NACVs. For example, it is not trivial to perform the direct fitting of the NACVs by the KRR method because the NACVs change dramatically and even diverge in the vicinity of the CI. In addition, the electronic structure calculations give random sign on electronic wave functions, resulting in an arbitrary sign for the NACVs. Some special tricks have to be used to overcome these problems.[61] This represents an interesting research topic in the future.

In summary, we presented a new approach to perform the nonadiabatic Zhu-Nakamura TSH dynamics simulation using preconstructed PESs based on a ML technique. The 6AP molecule was used as an example. After the proper sampling of the geometries in different important regions and the CASSCF calculations, the KRR method based on the CM descriptor was employed to obtain the PESs, in which adaptive and clustering strategies were also employed in the learning process. The trained ML-PESs display small MAEs (< 1 kcal/mol). Then, we employed the ML-PES in the TSH dynamics based on the Zhu-Nakamura algorithm. In the dynamics simulation, the PESs and gradients of most geometries were obtained by the ML-PESs, while we switched to the electronic structure calculations for only a few geometries that were either in the vicinity of the CIs or in the out-of-confidence region. For the 6AP model system, the ML-PES-based nonadiabatic dynamics simulation results, including the excited-state lifetime and reaction channels, were consistent with those obtained using the pure on-the-fly TSH simulation at the CASSCF levels. In addition, a massive dynamics simulation with 1000 trajectories was performed, which provided a much clearer picture of the dynamics process with a reasonable computational cost. This finding suggests that the inclusion of ML-PESs in the nonadiabatic dynamics simulation may allow the computation of a vast number of trajectories with a balance between computational efficiency and accuracy.

**Computational Details.** The GSM and vibrational normal modes of 6AP were obtained at the B3LYP/6-31G* level with the Gaussian 09 package,[85] which were further used to generate the initial conditions (geometries and velocities) of the trajectories based on the Wigner distribution. The SA-CASSCF method was used to obtain the adiabatic energies and gradients. To be consistent with previous study,[82] the SA-3-CASSCF(10,8) level was chosen in all of the CASSCF calculations of our work. The CASSCF calculations were performed with MOLPRO package.[86] The on-the-fly surface hopping dynamics simulations based on Zhu-Nakamura theory were performed in our work. Nuclear motion was propagated using the velocity-Verlet algorithm with a time step of 0.5 fs. Three adiabatic electronic states ($S_0$, $S_1$ and $S_2$) were involved in the dynamics processes according to previous study.[82] All of the dynamics simulations were performed in our developing version of the JADE package.[39-41] In this package, the module approach was used to combine the dynamics simulation with Fortran and the PES construction with Python. All ML-PES relevant codes, such as KRR, data prescreening, and the interface with dynamics, were written with the Python language, based on the scikit-learn toolkit.[87] The current work used CM[46-47,64,88] as the molecular descriptor. It is known that the CM is invariant to the translational and rotational motion of molecules, as well as the mirror reflection. However, the CM is not invariant to nuclear permutation. Although the reordering of the CM elements avoids this problem,[47] the resulting discontinuities in the input feature space may cause problems in the gradient calculations. Thus we still take the original CM as the input descriptor. However, the lack of permutation symmetry should not cause issues in the current work. First, the treatment of the nonadiabatic dynamics of 6AP does not require the consideration of the permutation. Although several identical atoms are found, they will not exchange their positions in the molecular dynamics. The only exception is that two hydrogen atoms in an $NH_2$ group may exchange their position by torsional motion. However, in excited-state dynamics, such torsional motion should not be very fast. Second, in many complex systems undergoing ultrafast nonadiabatic dynamics, such as the photostability of DNA, photoinduced proton transfer, and photoinduced electron/energy transfer without involving a large geometrical deformation, the permutation is also not relevant. In this situation, we can always use the CM as the molecular descriptor, just like the previous work.[64] Certainly, when the permutation symmetry is important, more advanced and sophisticated molecular descriptors, such as the symmetry functions,[49] permutation invariant polynomials,[89-90] local coordinate,[91] and bag-of-bonds[92] should be considered.

## AUTHOR INFORMATION

Corresponding Author

*Email: lanzg@qibebt.ac.cn; zhenggang.lan@gmail.com

**Notes**

The authors declare no competing financial interest.

## ACKNOWLEDGMENT

This work is supported by NSFC projects (21673266 and 21503248). This work was also supported by the Natural Science Foundation of Shandong Province for Distinguished Young Scholars (JQ201504). We also thank the Supercomputing Center, the Computer Network Information Center, CAS, and the Super Computational Center of CAS-QIBEBT for providing computational resources.

**Supporting Information Available**: Several theoretical methods and additional implementation details, including the Zhu-Nakamura method, molecular descriptors, confidence interval, kernel ridge regression, predictions of potential energies and gradients by the model from KRR, clustering, computational cost, and nonadiabatic dynamics results with on-the-fly Tully's surface-hoping method.

**REFERENCES**

(1) Domcke, W.; Yarkony, D. R.; Köppel, H. *Conical Intersections I: Electronic Structure, Dynamics and Spectroscopy*; World Scientific: Singapore; 2004.
(2) Domcke, W.; Yarkony, D. R.; Köppel, H. *Conical Intersections II: Theory, Computation and Experiment*; World Scientific: Singapore; 2011.
(3) Crespo-Hernández, C. E.; Cohen, B.; Hare, P. M.; Kohler, B. Ultrafast Excited-State Dynamics in Nucleic Acids. *Chem. Rev.* **2004**, *104*, 1977-2020.
(4) Middleton, C. T.; de La Harpe, K.; Su, C.; Law, Y. K.; Crespo-Hernández, C. E.; Kohler, B. DNA Excited-State Dynamics: from Single Bases to the Double Helix. *Annu. Rev. Phys. Chem.* **2009**, *60*, 217-39.
(5) Matsika, S.; Krause, P. Nonadiabatic Events and Conical Intersections. *Annu. Rev. Phys. Chem.* **2011**, *62*, 621-43.
(6) Levine, B. G.; Martínez, T. J. Isomerization through Conical Intersections. *Annu. Rev. Phys. Chem.* **2007**, *58*, 613-634.
(7) Gozem, S.; Luk, H. L.; Schapiro, I.; Olivucci, M. Theory and Simulation of the Ultrafast Double-Bond Isomerization of Biological Chromophores. *Chem. Rev.* **2017**, *117*, 13502-13565.
(8) Chergui, M. Ultrafast Photophysics of Transition Metal Complexes. *Acc. Chem. Res.* **2015**, *48*, 801-808.
(9) Eng, J.; Gourlaouen, C.; Gindensperger, E.; Daniel, C. Spin-Vibronic Quantum Dynamics for Ultrafast Excited-State Processes. *Acc. Chem. Res.* **2015**, *48*, 809-817.
(10) Meyer, H. D.; Manthe, U.; Cederbaum, L. S. The Multi-Configurational Time-Dependent Hartree Approach. *Chem. Phys. Lett.* **1990**, *165*, 73-78.
(11) Ben-Nun, M.; Martínez, T. J. Ab Initio Quantum Molecular Dynamics. *Adv. Chem. Phys.* **2002**, *121*, 439-512.
(12) Wang, H. B.; Thoss, M. Multilayer Formulation of the Multiconfiguration Time-Dependent Hartree Theory. *J. Chem. Phys.* **2003**, *119*, 1289-1299.
(13) Schwartz, B. J.; Bittner, E. R.; Prezhdo, O. V.; Rossky, P. J. Quantum Decoherence and the Isotope Effect in Condensed Phase Nonadiabatic Molecular Dynamics Simulations. *J. Chem. Phys.* **1996**, *104*, 5942-5955.
(14) Tully, J. C. Mixed Quantum-Classical Dynamics. *Faraday Discuss.* **1998**, *110*, 407-419.
(15) Sun, X.; Wang, H. B.; Miller, W. H. Semiclassical Theory of Electronically Nonadiabatic Dynamics: Results of a Linearized Approximation to the Initial Value Representation. *J. Chem. Phys.* **1998**, *109*, 7064-7074.
(16) Kapral, R. Progress in the Theory of Mixed Quantum-Classical Dynamics. *Annu. Rev. Phys. Chem.* **2006**, *57*, 129-157.
(17) Jasper, A. W.; Nangia, S.; Zhu, C. Y.; Truhlar, D. G. Non-Born-Oppenheimer Molecular Dynamics. *Acc. Chem. Res.* **2006**, *39*, 101-108.
(18) Abedi, A.; Maitra, N. T.; Gross, E. K. U. Exact Factorization of the Time-Dependent Electron-Nuclear Wave Function. *Phys. Rev. Lett.* **2010**, *105*, 123002.
(19) Yonehara, T.; Hanasaki, K.; Takatsuka, K. Fundamental Approaches to Nonadiabaticity: Toward a Chemical Theory beyond the Born-Oppenheimer Paradigm. *Chem. Rev.* **2012**, *112*, 499-542.

(20) Gorshkov, V. N.; Tretiak, S.; Mozyrsky, D. Semiclassical Monte-Carlo Approach for Modelling Non-Adiabatic Dynamics in Extended Molecules. *Nat. Commun.* **2013**, *4*, 2144.
(21) Wang, L. J.; Akimov, A.; Prezhdo, O. V. Recent Progress in Surface Hopping: 2011-2015. *J. Phys. Chem. Lett.* **2016**, *7*, 2100-2112.
(22) Tully, J. C. Molecular-Dynamics with Electronic-Transitions. *J. Chem. Phys.* **1990**, *93*, 1061-1071.
(23) Hammes-Schiffer, S.; Tully, J. C. Proton-Transfer in Solution - Molecular-Dynamics with Quantum Transitions. *J. Chem. Phys.* **1994**, *101*, 4657-4667.
(24) Belyaev, A. K.; Tiukanov, A. S.; Domcke, W. Generalized Diatomics-in-Molecules Method for Polyatomic Anions. *Phys. Rev. A* **2001**, *65*, 012508.
(25) Xie, W. W.; Domcke, W. Accuracy of Trajectory Surface-Hopping Methods: Test for a Two-Dimensional Model of the Photodissociation of Phenol. *J. Chem. Phys.* **2017**, *147*, 184114.
(26) Nikitin, E. E. e. *Theory of Elementary Atomic and Molecular Processes in Gases*; Oxford, Clarendon Press; 1974.
(27) Zhu, C.; Nakamura, H. Theory of Nonadiabatic Transition for General Two‐State Curve Crossing Problems. I. Nonadiabatic Tunneling Case. *J. Chem. Phys.* **1994**, *101*, 10630-10647.
(28) Zhu, C. Y.; Nakamura, H. Theory of Nonadiabatic Transition for General 2-State Curve Crossing Problems. II. Landau-Zener Case. *J. Chem. Phys.* **1995**, *102*, 7448-7461.
(29) Zhu, C. Y.; Nobusada, K.; Nakamura, H. New Implementation of the Trajectory Surface Hopping Method with Use of the Zhu-Nakamura Theory. *J. Chem. Phys.* **2001**, *115*, 3031-3044.
(30) Yu, L.; Xu, C.; Lei, Y. B.; Zhu, C. Y.; Wen, Z. Y. Trajectory-Based Nonadiabatic Molecular Dynamics without Calculating Nonadiabatic Coupling in the Avoided Crossing Case: Trans <-> Cis Photoisomerization in Azobenzene. *Phys. Chem. Chem. Phys.* **2014**, *16*, 25883-25895.
(31) Yue, L.; Yu, L.; Xu, C.; Lei, Y.; Liu, Y.; Zhu, C. Benchmark Performance of Global Switching versus Local Switching for Trajectory Surface Hopping Molecular Dynamics Simulation: Cis<-->Trans Azobenzene Photoisomerization. *ChemPhysChem* **2017**, *18*, 1274-1287.
(32) Persico, M.; Granucci, G. An Overview of Nonadiabatic Dynamics Simulations Methods, with Focus on the Direct Approach Versus the Fitting of Potential Energy Surfaces. *Theor. Chem. Acc.* **2014**, *133*, 1526.
(33) Doltsinis, N. L.; Marx, D. Nonadiabatic Car-Parrinello Molecular Dynamics. *Phys. Rev. Lett.* **2002**, *88*, 166402.
(34) Barbatti, M.; Granucci, G.; Persico, M.; Ruckenbauer, M.; Vazdar, M.; Eckert-Maksić, M.; Lischka, H. The On-the-Fly Surface-Hopping Program System NEWTON-X: Application to Ab Initio Simulation of the Nonadiabatic Photodynamics of Benchmark Systems. *J. Photochem. Photobiol. A* **2007**, *190*, 228-240.
(35) Fabiano, E.; Thiel, W. Nonradiative Deexcitation Dynamics of 9H-Adenine: An OM2 Surface Hopping Study. *J. Phys. Chem. A* **2008**, *112*, 6859-6863.
(36) Werner, U.; Mitrić, R.; Suzuki, T.; Bonačić-Koutecký, V. Nonadiabatic Dynamics within the Time Dependent Density Functional Theory: Ultrafast Photodynamics in Pyrazine. *Chem. Phys.* **2008**, *349*, 319-324.

(37) Tavernelli, I.; Tapavicza, E.; Rothlisberger, U. Non-Adiabatic Dynamics Using Time-Dependent Density Functional Theory: Assessing the Coupling Strengths. *J. Mol. Struct.* **2009**, *914*, 22-29.

(38) Richter, M.; Marquetand, P.; González-Vázquez, J.; Sola, I.; González, L. SHARC: Ab Initio Molecular Dynamics with Surface Hopping in the Adiabatic Representation Including Arbitrary Couplings. *J. Chem. Theory Comput.* **2011**, *7*, 1253-1258.

(39) Du, L.; Lan, Z. An On-the-Fly Surface-Hopping Program JADE for Nonadiabatic Molecular Dynamics of Polyatomic Systems: Implementation and Applications. *J. Chem. Theory Comput.* **2015**, *11*, 1360-1374.

(40) Du, L. K.; Lan, Z. G. An On-the-Fly Surface-Hopping Program JADE for Nonadiabatic Molecular Dynamics of Polyatomic Systems: Implementation and Applications (vol 11, pg 1360, 2015). *J. Chem. Theory Comput.* **2015**, *11*, 4522-4523.

(41) Hu, D. P.; Liu, Y. F.; Sobolewski, A. L.; Lan, Z. G. Nonadiabatic Dynamics Simulation of Keto Isocytosine: A Comparison of Dynamical Performance of Different Electronic-Structure Methods. *Phys. Chem. Chem. Phys.* **2017**, *19*, 19168-19177.

(42) Bishop, C. *Pattern Recognition and Machine Learning*; Springer-Verlag: New York; 2006.

(43) Raff, L. *Neural Networks in Chemical Reaction Dynamics*; Oxford University Press: New York; 2012.

(44) Ramakrishnan, R.; von Lilienfeld, O. A. Machine Learning, Quantum Chemistry, and Chemical Space. *Rev. Comput. Chem.* **2017**, *30*, 225-256.

(45) Pyzer-Knapp, E. O.; Li, K.; Aspuru-Guzik, A. Learning from the Harvard Clean Energy Project: The Use of Neural Networks to Accelerate Materials Discovery. *Adv. Funct. Mater.* **2015**, *25*, 6495-6502.

(46) Rupp, M.; Ramakrishnan, R.; von Lilienfeld, O. A. Machine Learning for Quantum Mechanical Properties of Atoms in Molecules. *J. Phys. Chem. Lett.* **2015**, *6*, 3309-3313.

(47) Rupp, M. Machine Learning for Quantum Mechanics in a Nutshell. *Int. J. Quantum Chem.* **2015**, *115*, 1058-1073.

(48) Bartók, A. P.; De, S.; Poelking, C.; Bernstein, N.; Kermode, J. R.; Csányi, G.; Ceriotti, M. Machine Learning Unifies the Modeling of Materials and Molecules. *Sci. Adv.* **2017**, *3*, e1701816.

(49) Behler, J.; Parrinello, M. Generalized Neural-Network Representation of High-Dimensional Potential-Energy Surfaces. *Phys. Rev. Lett.* **2007**, *98*, 146401.

(50) Chen, J.; Xu, X.; Xu, X.; Zhang, D. H. Communication: An Accurate Global Potential Energy Surface for the OH + CO -> H + CO2 Reaction Using Neural Networks. *J. Chem. Phys.* **2013**, *138*, 221104.

(51) Behler, J. Constructing High-Dimensional Neural Network Potentials: A Tutorial Review. *Int. J. Quantum Chem.* **2015**, *115*, 1032-1050.

(52) Li, Z. W.; Kermode, J. R.; De Vita, A. Molecular Dynamics with On-the-Fly Machine Learning of Quantum-Mechanical Forces. *Phys. Rev. Lett.* **2015**, *114*, 096405.

(53) Botu, V.; Ramprasad, R. Adaptive Machine Learning Framework to Accelerate Ab Initio Molecular Dynamics. *Int. J. Quantum Chem.* **2015**, *115*, 1074-1083.

(54) Gastegger, M.; Marquetand, P. High-Dimensional Neural Network Potentials for Organic Reactions and an Improved Training Algorithm. *J. Chem. Theory Comput.* **2015**, *11*, 2187-2198.

(55) Behler, J. Perspective: Machine Learning Potentials for Atomistic Simulations. *J. Chem. Phys.* **2016**, *145*, 170901.
(56) Jiang, B.; Li, J.; Guo, H. Potential Energy Surfaces from High Fidelity Fitting of Ab Initio Points: the Permutation Invariant Polynomial - Neural Network Approach. *Int. Rev. Phys. Chem.* **2016**, *35*, 479-506.
(57) Shao, K.; Chen, J.; Zhao, Z.; Zhang, D. H. Communication: Fitting Potential Energy Surfaces with Fundamental Invariant Neural Network. *J. Chem. Phys.* **2016**, *145*, 071101.
(58) Shen, L.; Wu, J. H.; Yang, W. T. Multiscale Quantum Mechanics/Molecular Mechanics Simulations with Neural Networks. *J. Chem. Theory Comput.* **2016**, *12*, 4934-4946.
(59) Chmiela, S.; Tkatchenko, A.; Sauceda, H. E.; Poltavsky, I.; Schuett, K. T.; Müeller, K.-R. Machine Learning of Accurate Energy-Conserving Molecular Force Fields. *Sci. Adv.* **2017**, *3*, e1603015.
(60) Richings, G. W.; Habershon, S. Direct Quantum Dynamics Using Grid-Based Wave Function Propagation and Machine-Learned Potential Energy Surfaces. *J. Chem. Theory Comput.* **2017**, *13*, 4012-4024.
(61) Guan, Y.; Fu, B.; Zhang, D. H. Construction of Diabatic Energy Surfaces for Lifh with Artificial Neural Networks. *J. Chem. Phys.* **2017**, *147*, 224307.
(62) Gastegger, M.; Behler, J.; Marquetand, P. Machine Learning Molecular Dynamics for the Simulation of Infrared Spectra. *Chem. Sci.* **2017**, *8*, 6924-6935.
(63) Shen, L.; Yang, W. T. Molecular Dynamics Simulations with Quantum Mechanics/Molecular Mechanics and Adaptive Neural Networks. *J. Chem. Theory Comput.* **2018**, *14*, 1442-1455.
(64) Häse, F.; Valleau, S.; Pyzer-Knapp, E.; Aspuru-Guzik, A. Machine Learning Exciton Dynamics. *Chem. Sci.* **2016**, *7*, 5139-5147.
(65) Ferré, G.; Haut, T.; Barros, K. Learning Molecular Energies Using Localized Graph Kernels. *J. Chem. Phys.* **2017**, *146*, 114107.
(66) Dral, P. O.; Owens, A.; Yurchenko, S. N.; Thiel, W. Structure-Based Sampling and Self-Correcting Machine Learning for Accurate Calculations of Potential Energy Surfaces and Vibrational Levels. *J. Chem. Phys.* **2017**, *146*, 244108.
(67) Unke, O. T.; Meuwly, M. Toolkit for the Construction of Reproducing Kernel-Based Representations of Data: Application to Multidimensional Potential Energy Surfaces. *J. Chem. Inf. Model.* **2017**, *57*, 1923-1931.
(68) Manzhos, S.; Carrington Jr., T. A Random-Sampling High Dimensional Model Representation Neural Network for Building Potential Energy Surfaces. *J. Chem. Phys.* **2006**, *125*, 084109.
(69) Pukrittayakamee, A.; Malshe, M.; Hagan, M.; Raff, L. M.; Narulkar, R.; Bukkapatnum, S.; Komanduri, R. Simultaneous Fitting of a Potential-Energy Surface and Its Corresponding Force Fields Using Feedforward Neural Networks. *J. Chem. Phys.* **2009**, *130*, 134101.
(70) Nguyen-Truong, H. T.; Le, H. M. An Implementation of the Levenberg–Marquardt Algorithm for Simultaneous-Energy-Gradient Fitting Using Two-Layer Feed-Forward Neural Networks. *Chem. Phys. Lett.* **2015**, *629*, 40-45.
(71) Cubuk, E. D.; Malone, B. D.; Onat, B.; Waterland, A.; Kaxiras, E. Representations in Neural Network Based Empirical Potentials. *J. Chem. Phys.* **2017**, *147*, 024104.

(72) Kolb, B.; Luo, X.; Zhou, X.; Jiang, B.; Guo, H. High-Dimensional Atomistic Neural Network Potentials for Molecule-Surface Interactions: HCl Scattering from Au(111). *J. Phys. Chem. Lett.* **2017**, *8*, 666-672.

(73) Lenzen, T.; Manthe, U. Neural Network Based Coupled Diabatic Potential Energy Surfaces for Reactive Scattering. *J. Chem. Phys.* **2017**, *147*, 084105.

(74) Schütt, K. T.; Arbabzadah, F.; Chmiela, S.; Müller, K. R.; Tkatchenko, A. Quantum-Chemical Insights from Deep Tensor Neural Networks. *Nat. Commun.* **2017**, *8*, 13890.

(75) Smith, J. S.; Isayev, O.; Roitberg, A. E. ANI-1: An Extensible Neural Network Potential with DFT Accuracy at Force Field Computational Cost. *Chem. Sci.* **2017**, *8*, 3192-3203.

(76) Bartók, A. P.; Payne, M. C.; Kondor, R.; Csányi, G. Gaussian Approximation Potentials: The Accuracy of Quantum Mechanics, without the Electrons. *Phys. Rev. Lett.* **2010**, *104*, 136403.

(77) Agrawal, P. M.; Raff, L. M.; Hagan, M. T.; Komanduri, R. Molecular Dynamics Investigations of the Dissociation of $SiO_2$ on an Ab Initio Potential Energy Surface Obtained Using Neural Network Methods. *J. Chem. Phys.* **2006**, *124*, 134306.

(78) Zhu, X. L.; Yarkony, D. R. Fitting Coupled Potential Energy Surfaces for Large Systems: Method and Construction of a 3-State Representation for Phenol Photodissociation in the Full 33 Internal Degrees of Freedom Using Multireference Configuration Interaction Determined Data. *J. Chem. Phys.* **2014**, *140*, 024112.

(79) Malbon, C. L.; Yarkony, D. R. Nonadiabatic Photodissociation of the Hydroxymethyl Radical from the 22A State. Surface Hopping Simulations Based on a Full Nine-Dimensional Representation of the 1,2,32A Potential Energy Surfaces Coupled by Conical Intersections. *J. Phys. Chem. A* **2015**, *119*, 7498-7509.

(80) Zhu, X.; Malbon, C. L.; Yarkony, D. R. An Improved Quasi-Diabatic Representation of the 1, 2, $3^1$ A Coupled Adiabatic Potential Energy Surfaces of Phenol in the Full 33 Internal Coordinates. *J. Chem. Phys.* **2016**, *144*, 124312.

(81) Zhu, X.; Yarkony, D. R. On the Elimination of the Electronic Structure Bottleneck in on the Fly Nonadiabatic Dynamics for Small to Moderate Sized (10-15 Atom) Molecules Using Fit Diabatic Representations Based Solely on Ab Initio Electronic Structure Data: The Photodissociation of Phenol. *J. Chem. Phys.* **2016**, *144*, 024105.

(82) Barbatti, M.; Lischka, H. Can the Nonadiabatic Photodynamics of Aminopyrimidine Be a Model for the Ultrafast Deactivation of Adenine? *J. Phys. Chem. A* **2007**, *111*, 2852-2858.

(83) Müller, U.; Stock, G. Surface-Hopping Modeling of Photoinduced Relaxation Dynamics on Coupled Potential-Energy Surfaces. *J. Chem. Phys.* **1997**, *107*, 6230-6245.

(84) Weingart, O.; Lan, Z.; Koslowski, A.; Thiel, W. Chiral Pathways and Periodic Decay in cis-Azobenzene Photodynamics. *J. Phys. Chem. Lett.* **2011**, *2*, 1506-1509.

(85) Frisch, M. J.; Trucks, G. W.; Schlegel, H. B.; Scuseria, G. E.; Robb, M. A.; Cheeseman, J. R.; Scalmani, G.; Barone, V.; Mennucci, B.; Petersson, G. A.*; et al. Gaussian 09*; Gaussian, Inc.: Wallingford, CT, USA; 2009.

(86) *MOLPRO, version 2012.1, A Package of Ab Initio Programs, H.-J. Werner, P. J. Knowles, G. Knizia, F. R. Manby, M. Schütz, and others , see* http://www.molpro.net.

(87) Pedregosa, F.; Varoquaux, G.; Gramfort, A.; Michel, V.; Thirion, B.; Grisel, O.; Blondel, M.; Prettenhofer, P.; Weiss, R.; Dubourg, V. Scikit-Learn: Machine learning in Python. *J. Mach. Learn. Res.* **2011**, *12*, 2825-2830.

(88) Rupp, M.; Tkatchenko, A.; Muller, K. R.; von Lilienfeld, O. A. Fast and Accurate Modeling of Molecular Atomization Energies with Machine Learning. *Phys. Rev. Lett.* **2012**, *108*, 058301.
(89) Braams, B. J.; Bowman, J. M. Permutationally Invariant Potential Energy Surfaces in High Dimensionality. *Int. Rev. Phys. Chem.* **2009**, *28*, 577-606.
(90) Bowman, J. M.; Czakó, G.; Fu, B. High-Dimensional Ab Initio Potential Energy Surfaces for Reaction Dynamics Calculations. *Phys. Chem. Chem. Phys.* **2011**, *13*, 8094-8111.
(91) Zhang, L. F.; Han, J. Q.; Wang, H.; Car, R.; Weinan, E. Deep Potential Molecular Dynamics: A Scalable Model with the Accuracy of Quantum Mechanics. *Phys. Rev. Lett.* **2018**, *120*, 143001.
(92) Hansen, K.; Biegler, F.; Ramakrishnan, R.; Pronobis, W.; von Lilienfeld, O. A.; Müller, K. R.; Tkatchenko, A. Machine Learning Predictions of Molecular Properties: Accurate Many-Body Potentials and Nonlocality in Chemical Space. *J. Phys. Chem. Lett.* **2015**, *6*, 2326-2331.

Supporting information for

# Inclusion of Machine Learning Kernel Ridge Regression Potential Energy Surfaces in On-the-Fly Nonadiabatic Molecular Dynamics Simulation

*Deping Hu,[†,‡] Yu Xie,[†] Xusong Li,[†,‡] Lingyue Li,[‡] and Zhenggang Lan\*[†,‡]*

[†]CAS Key Laboratory of Biobased Materials, Qingdao Institute of Bioenergy and Bioprocess Technology, Chinese Academy of Sciences, Qingdao 266101, China

[‡]University of Chinese Academy of Sciences, Beijing 100049, China

Email: lanzg@qibebt.ac.cn, zhenggang.lan@gmail.com

**S1. Zhu-Nakamura method**

In trajectory surface hopping (TSH) dynamics simulation, the trajectories are allowed to hop between different adiabatic excited states to describe nonadiabatic transitions. Zhu and Nakamura once proposed a global switching algorithm to compute the hopping probability. Recently, Zhu and co-workers tried to make the further approximation of the Zhu-Nakamura formulism, which allows the estimation of hopping probability without the calculations of nonadiabatic coupling vectors (NACVs).[1-3] Because the detailed explanation of this algorithm is not the main purpose of this work, we suggest the interesting reader to check all details in previous works.[1-3]

This simplified approach achieves great success in nonadiabatic dynamics simulation of several systems.[2-3] Previous study made a detailed comparison between Zhu-Nakamura method and Tully's fewest switches method, which shows excellent agreement between the results obtained by two hopping algorithms.[3] Thus, the TSH based on the Zhu-Nakamura simplified algorithm was used in our work, because the NACV calculations are not involved.

**S2. Molecular descriptors**

In the potential energy surface (PES) construction process, one of the critical issues is how to define proper molecular descriptors to represent different geometries.

Although the molecular geometry can be naturally represented by Cartesian coordinates, such geometrical descriptor is not a good choice to represent the intramolecular motions. Normally it is possible to construct the molecular descriptors by the nonlinear transformation of nuclear Cartesian coordinates, However, such transformation is not uniquely defined, resulting in various molecular descriptor.

One simple molecular descriptor is the so-called Coulomb matrix (CM),[4] namely,

$$M_{kl} = \begin{cases} 0.5 Z_k^{2.4} & k = l \\ \dfrac{Z_k Z_l}{\|R_k - R_l\|} & k \neq l \end{cases} \quad (1)$$

where $Z_k$ is the atomic number of atom $k$, $R_k$ is the Cartesian position vector of atom $k$ and $\|R_k\text{-}R_l\|$ is the Euclidean distance between atom $k$ and $l$. The CM provides a simple and effective representation of molecular structures, which takes both element types and internal distances into account. Rupp, Lilienfeld and their co-workers employed this descriptor to predict molecular properties.[4] Aspuru-Guzik and co-workers used this descriptor in the construction of some elements in the diabatic Hamilton of photoharvesting systems.[5] Thus, in principle the CM can be used as the input feature in the PES construction. In the current work, we performed the normalization over the training data for each elements of CM matrix, because our results show that such normalization gives a slightly better machine learning PES (ML-PES). For same isolated system, the diagonal element of CM remains unchanged with respect to different geometries. Thus, after removing these elements and the

normalization of dataset, the employment of CM is equivalent to the use of the normalized inversed distance matrix in the PES fitting. We noticed that previous works once used distance matrix or inversed one as the input descriptor in PES fitting.[6] Next, we could like to make several comments on the employment of CM.

To treat the system-environment system, the CM should include the contribution of environment.[5] Then the diagonal elements changes as the functions of all coordinates and thus their terms cannot be neglected in PES construction. In such case, the CM and the distance matrix are not equivalent, even after normalization.

It is known that the CM is invariant to the translational and rotational motion of the molecule, as well as the mirror reflection. However, the CM is not invariant to nuclear permutation. Previous work introduced the re-order of matrix elements to avoid this problem, while this way introduce other problems, such as discontinuities.[7] Because the discontinuities in the input feature space may cause problems in the gradient calculations from the fitting PES, this is not a good approach for the molecular dynamics study. Considering this situation, we still take the original CM as the input descriptor. This means that our current approaches cannot describe the permutation symmetry of molecular system. However, this does not cause any trouble in the current work. First, the treatment of the nonadiabatic dynamics of the current model system, 6AP, does not require the consideration of permutation. Although several identical atoms are founded, they will not exchange their position in molecular dynamics. The only exception is that two hydrogen atoms in $NH_2$ group may exchange their position by torsional motion. However, in excited state dynamics, such torsional motion should not be very fast. Thus, we still can assume that the interchanges of two hydrogen atoms are not relevant to ultrafast nonadiabatic dynamics. This means that the permutation symmetry is not very important for the current system after proper sampling. Second, in many complex systems undergoing ultrafast nonadiabatic dynamics, such as the photostability of DNA, photoinduced proton transfer, photoinduced electron/energy transfer without involving large geometry deformation, the permutation is also not relevant. In this situation, we can always use the CM as the molecular descriptor, just like the previous work.[5] Certainly, when the permutation symmetry is un-avoided, more advanced and sophisticated molecular descriptors, such as the symmetry functions,[8] permutation invariant polynomial,[9-10] local coordinate,[11] and bag-of-bonds[12] should be considered. The employment of them requires additional efforts because their construction is not trivial.

**S3. Confidence interval**

In the current work, the CM of all geometries in the training dataset were employed to describe the sampling space. After the off-diagonal elements of the upper triangle matrix were taken, each molecular descriptor becomes a vector with multiple dimensions. Here we employed an approximate strategy to compute the confidence interval in the sampling space covered by the CMs of all geometries in the training set.

For each dimension, we start from a trial value x% (for example 99%) to define a one-dimensional confidence interval. This means that x% of all geometries in the training data set is in this one-dimensional confidence interval for a particular dimension. The x% value is the same for each dimension. This step is easily done by using Scikit-learn package,[13] which also provides the reliable region $[a_{min}^{(i)}, a_{max}^{(i)}]$ for all dimensions. After we repeat this examination over all dimensions, we assume that a geometry (in the training set) locates in the multi-dimensional confidence interval spanned by all CMs, only if such geometry stays in the one-dimensional confidence intervals for all dimensions. Next the number of geometries in this multi-dimensional confidence interval are calculated and we divide this number by the total number of geometries in the training set to get the percentage (y%). Approximately, this means that the multi-dimensional confidence interval covers y% of total geometries in the training set. Next, we adjust x% until y=95 is reached. This means 95% of geometries in training sets are in the multi-dimensional confidence interval by using the optimal value x% for each dimension.

After the optimal value x% and the reliable region $[a_{min}^{(i)}, a_{max}^{(i)}]$ are determined for each dimension, it is possible to know whether any new geometry is in confidence interval or not. We compute the CM of any given geometry in the dynamics. If the corresponding element of CM for the *i*-th dimension is not in $[a_{min}^{(i)}, a_{max}^{(i)}]$, this geometry is in out-of-confidence regions.

**S4. Kernel ridge regression**

Kernel ridge regression (KRR) is one of the most used supervised learning approaches, which has been used for prediction of molecular properties in several studies.[4,7] The basic idea of KRR is to map input features into a higher-dimensional space, where linear relation between the transformed features and the property of interest could be established. The kernel trick that depends on the inner product of the input features is often used in the mapping process.

In the KRR approach, molecular property $f(\boldsymbol{m})$ is estimated by the function of a nuclear configuration with molecular descriptor $\boldsymbol{m}$ could be calculated as the following expression,

$$f(\boldsymbol{m}) = \sum_{j=1}^{N_t} c_j \mathrm{K}(\boldsymbol{m}, \boldsymbol{m}^{(j)}) \qquad (2)$$

where $\boldsymbol{m}^{(j)}$ is the molecular descriptor for the *j*-th configuration and $N_t$ is the number of configurations in the training set, $c_j$ is the regression coefficients. K is the kernel function, in our work, the radial basis function (RBF) kernel is used in our learning algorithms. The RBF kernel is often called Gaussian kernel defined as:

$$\mathrm{K}(\boldsymbol{m}, \boldsymbol{m}^{(j)}) = \exp\left(-\frac{\|\boldsymbol{m} - \boldsymbol{m}^{(j)}\|^2}{2\sigma^2}\right) \qquad (3)$$

where $\sigma$ is the kernel width, $\|\boldsymbol{m} - \boldsymbol{m}^{(j)}\|$ is the Euclidean distance between the two molecular descriptor vectors $\boldsymbol{m}$ and $\boldsymbol{m}^{(j)}$. The regression coefficients $c_j$ is trained by minimizing the following expression:

$$\sum_{j=1}^{N_t}[f(\boldsymbol{m}^{(j)}) - f^{ref}(\boldsymbol{m}^{(j)})] + \lambda \boldsymbol{C}^T \boldsymbol{K} \boldsymbol{C} \tag{4}$$

where $f^{ref}$ is reference molecular property value, the kernel matrix **K** is obtained by Equation (3) over all pairwise distances between all training data, $\lambda$ is a regularization parameter, which is used to prevent overfitting.

**S5. Predictions of potential energies and gradients by the model from KRR.**

Below we give the formal derivation of potential and gradient when the KRR model is constructed. The molecular geometry can be defined in the Cartesian space with $3 \times Na$ dimensionality, where $Na$ is the number of atom.

At a given geometry it is always possible to compute the corresponding CM by using Equation (1). Please notice that diagonal elements of CM are not dependent on configuration, and thus it is safe to remove them in the construction of molecular descriptor. In addition, the CM is symmetric, thus we only need to consider the off-diagonal elements of the upper triangle part. All these used off-diagonal elements define a vector *M* that contains $Na \times (Na-1)/2$ elements. Each element of *M* is labeled as $M_s$. After the normalization of each dimension of *M* over all training geometries, we get another vector $\boldsymbol{m} = \boldsymbol{W}\boldsymbol{M}$, where *W* is a diagonal matrix used mainly for normalization. This transformation gives: $m_s = W_s M_s$.

The *M* vector of the *i*-th geometry in the training set is labeled as $\boldsymbol{M}^{(i)}$, and the $M_s^{(i)}$ is the component of this vector in the $s$-th dimension. $W_s$ is the factor obtained from the normalization of $M_s^{(i)}$ over all training geometries. Please notice that such factor becomes 1 for some dimension if no normalization is performed.

Overall, at any geometry, it is always possible to calculate the corresponding $\boldsymbol{m}$ (namely all $m_s$) from the CM and original Cartesian coordinates, see Scheme S1.

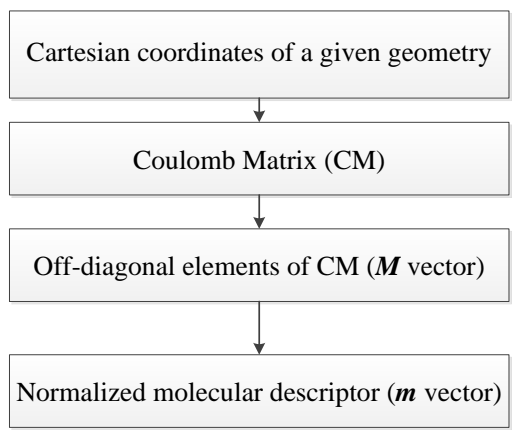

**Scheme S1**. The construction of molecular descriptor.

The molecular property $f(\boldsymbol{m})$ (for instance potential energy of an electronic state) is calculated using following expression:

$$f(\boldsymbol{m}) = \sum_{i=1}^{N_t} c_i K(\boldsymbol{m}, \boldsymbol{m}^{(i)}) \quad (5)$$

where the summation is over all training geometries and the Gaussian kernel is employed:

$$K(\boldsymbol{m}, \boldsymbol{m}^{(i)}) = \exp(-\frac{\|\boldsymbol{m} - \boldsymbol{m}^{(i)}\|^2}{2\sigma^2}) \quad (6)$$

As discussed previously, the $c_i$ and $\sigma$ can be obtained by the training/validation steps. Because the potential energies can be expressed in the above equation, it is rather trivial to obtain the potential energy for a new geometry.

The gradient is the first-order derivative of adiabatic energy $f(\boldsymbol{m})$ with respect to Cartesian coordinates $(x, y, z)$ of each atom. Let us take the coordinate $x_k$ as an example, which refers to the $x$ coordinate of atom $k$, the gradient with respect to $x_k$ is calculated by the chain rule:

$$\frac{\partial f(\boldsymbol{m})}{\partial x_k} = \sum_s \frac{\partial f(\boldsymbol{m})}{\partial m_s} \frac{\partial m_s}{\partial x_k} \qquad (7)$$

Here, the $m_s$ is an element of $\boldsymbol{m}$. After the construction of $f(\boldsymbol{m})$, it is always possible to get the derivative $\frac{\partial f(\boldsymbol{m})}{\partial m_s}$. According to Scheme S1 and above discussions, $m_s = W_s M_s$ and $M_s$ is just an element of CM that is obtained from Cartesian coordinate of each atom according to Equation (1). Because all analytical expressions are easily constructed, it is possible to get derivatives.

In the current work, we employed the PES data in the fitting procedure of the KRR methods. When the PES functions are constructed, it is rather easy to obtain the gradient of the adiabatic states.

In principle, it is also possible to include the gradients in the KRR fitting procedure, while additional numerical tricks should be considered in the regression steps.

If we consider the gradients as independent physical properties in the KRR step,

namely fitting both of PES and gradient separately, the totally energy may not be conserved in the dynamics simulation.[6]

Alternatively, it is also possible to perform the KRR fitting of PES and gradient in a correlated way. This means that we use the fitting function with Gaussian kernels to define the PES and then derive the gradient expression. Lately, we fit both PES and gradients according to the same set of Gaussian kernels and parameters.[6] This way requires some additional numerical tricks. For instance, the overall regression deviation is composed of PES and gradient fitting errors, while a penalty function is employed to adjust their ratio. It is highly recommended to use some tricks to determine this ratio.[11,14-15]

In the current work, we take a rather simple approach to only fit PES and calculate gradient afterwards. This approach gives us a reasonable result and thus such approach is reliable. In the future work, it is certainly interesting to include the gradient in training step as well, because this may allow us to reduce the number of points in the training set.[16-17]

**S6. Clustering**

After the geometry sampling, a large number of geometries are generated. When we directly use all of them to perform the regression, two critical issues exist. First, because a huge number of data points are used in the KRR regression, the fitting procedure may become extremely slow. Sometimes the regression may require a huge amount of computer memory that is beyond the hardware status of many widely-used computers. Second, sampling methods may provide several points that are located within the same areas, while much less points may appear in some critical areas or other useful areas. Thus, it may be useful to create a balanced description on the data distribution over all important space in the regression, as suggested by previous work.[5]

One idea is to perform the clustering analysis of data points before training. The similarity between CM of different geometries is taken as the reference to perform clustering with the hierarchical agglomerative clustering algorithm.[18] After all geometries are grouped into many clusters, we randomly pick up the same number of geometries from each cluster, and combine them to form a new dataset for the next training step.

Because the clustering step is employed to reduce the computational cost and get balanced geometry distribution, we mainly focus on the final fitting accuracy. In the current work, we select 700 clusters and the good fitting results are given. In some sense, this clustering trick is only for practical reasons and the geometrical feature in each cluster is beyond the scope of the current discussion.

Certainly, we still can say that we find some "features" in the distribution of geometry numbers in each cluster. To explain this situation, we present the distribution of the 6AP conformation number in each cluster. As shown in Figure S1, the distribution is neither uniform nor Gaussian, in which many groups have more than 300 6AP conformations, while some have less than 50 6AP conformations. We noticed that the maximum value of this distribution appears near 100. In this case, we only naively select 100 geometries from each cluster. When a cluster contains the geometries less than 100, we will select all of them.

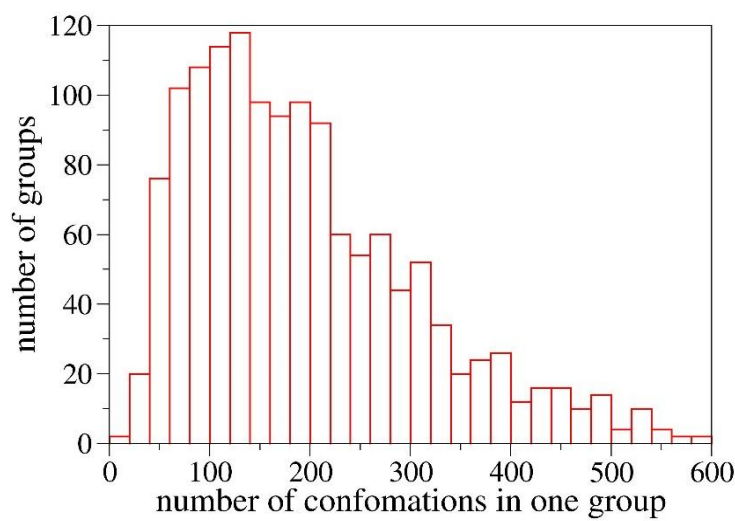

**Figure S1.** Distribution of the number of 6AP conformations for each group clustered using the hierarchical agglomerative clustering algorithm for the training set.

**S7. Computational cost**

In nonadiabatic dynamics simulation, most computational cost comes from the electronic structure calculations. For one geometry of 6AP, we need ~90 seconds [in the computer with Intel(R) Xeon(R) CPU E3-1230 v3 @ 3.30GHz] to calculate the energies and gradients at SA3-CASSCF(10,8) level using MOLPRO package and no parallel calculation is involved. As a contrast, we just need ~5 seconds to predict them using ML-PESs in the same computing environment. It takes ~50 hours for calculation of a trajectory with 2000 steps in the pure on-the-fly calculation. Much less time ~4.7 hours is required, if the dynamics mainly uses ML-PESs and only ~4% geometries are treated by CASSCF calculations.

We have to say that the current work represents our initial efforts to employ the ML-PES in nonadiabatic dynamics. The coding effort largely relies on our developing nonadiabatic dynamics code JADE, in which a good interface between the CASSCF calculations and dynamics calculation was developed.[19] This allows us easily to switch between ML-PES and CASSCF calculations. Thus, we load the KRR model only when we need to do so at each step. Because the large number of data are read before the prediction of potential energy and gradients, the I/O time is rather long. Thus, the above time is not the fully optimal for such trajectory calculations. However, the employment of ML-PES is certainly much faster than the direct electronic structure calculations.

**S8. Nonadiabatic dynamics simulation with on-the-fly Tully's surface-hoping method**

To illustrate that Zhu-Nakamura theory is reliable for dynamics simulation on 6AP, the on-the-fly Tully's TSH dynamics simulation at the same CASSCF level is also performed for benchmark, see Figure S2. The nuclear motion was propagated using the velocity-Verlet algorithm with a time step of 0.5 fs and the electronic motion was propagated with a time step of 0.005 fs. In this system, the dynamics results, including excited-state lifetime and reaction channels, consist with the results based on Zhu-Nakamura theory.

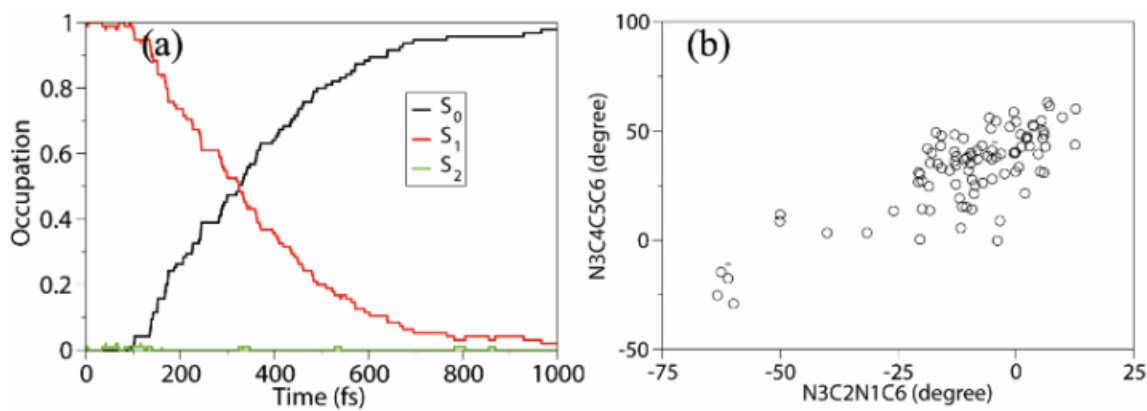

**Figure S2.** Tully's TSH nonadiabatic dynamics results with 100 trajectories initiated from $S_1$ at CASSCF level: (a) Time-dependent average fractional occupations of the adiabatic electronic states; and (b) the geometry distributions of dihedral N3C2N1C6 and N3C4C5C6 at hops in the simulation process.

# References

(1) Zhu, C. Y.; Nobusada, K.; Nakamura, H. New Implementation of the Trajectory Surface Hopping Method with Use of the Zhu-Nakamura Theory. *J. Chem. Phys.* **2001**, *115*, 3031-3044.
(2) Yu, L.; Xu, C.; Lei, Y. B.; Zhu, C. Y.; Wen, Z. Y. Trajectory-Based Nonadiabatic Molecular Dynamics without Calculating Nonadiabatic Coupling in the Avoided Crossing Case: Trans <-> Cis Photoisomerization in Azobenzene. *Phys. Chem. Chem. Phys.* **2014**, *16*, 25883-25895.
(3) Yue, L.; Yu, L.; Xu, C.; Lei, Y.; Liu, Y.; Zhu, C. Benchmark Performance of Global Switching versus Local Switching for Trajectory Surface Hopping Molecular Dynamics Simulation: Cis<-->Trans Azobenzene Photoisomerization. *Chemphyschem* **2017**, *18*, 1274-1287.
(4) Rupp, M.; Tkatchenko, A.; Muller, K. R.; von Lilienfeld, O. A. Fast and Accurate Modeling of Molecular Atomization Energies with Machine Learning. *Phys. Rev. Lett.* **2012**, *108*, 058301.
(5) Häse, F.; Valleau, S.; Pyzer-Knapp, E.; Aspuru-Guzik, A. Machine Learning

Exciton Dynamics. *Chem. Sci.* **2016**, *7*, 5139-5147.
(6) Raff, L. *Neural Networks in Chemical Reaction Dynamics*; Oxford University Press: New York; 2012.
(7) Rupp, M. Machine Learning for Quantum Mechanics in a Nutshell. *Int. J. Quantum Chem.* **2015**, *115*, 1058-1073.
(8) Behler, J.; Parrinello, M. Generalized Neural-Network Representation of High-Dimensional Potential-Energy Surfaces. *Phys. Rev. Lett.* **2007**, *98*, 146401.
(9) Braams, B. J.; Bowman, J. M. Permutationally Invariant Potential Energy Surfaces in High Dimensionality. *Int. Rev. Phys. Chem.* **2009**, *28*, 577-606.
(10) Bowman, J. M.; Czakó, G.; Fu, B. High-Dimensional Ab Initio Potential Energy Surfaces for Reaction Dynamics Calculations. *Phys. Chem. Chem. Phys.* **2011**, *13*, 8094-8111.
(11) Zhang, L. F.; Han, J. Q.; Wang, H.; Car, R.; Weinan, E. Deep Potential Molecular Dynamics: A Scalable Model with the Accuracy of Quantum Mechanics. *Phys. Rev. Lett.* **2018**, *120*, 143001.
(12) Hansen, K.; Biegler, F.; Ramakrishnan, R.; Pronobis, W.; von Lilienfeld, O. A.; Müller, K. R.; Tkatchenko, A. Machine Learning Predictions of Molecular Properties: Accurate Many-Body Potentials and Nonlocality in Chemical Space. *J. Phys. Chem. Lett.* **2015**, *6*, 2326-2331.
(13) Pedregosa, F.; Varoquaux, G.; Gramfort, A.; Michel, V.; Thirion, B.; Grisel, O.; Blondel, M.; Prettenhofer, P.; Weiss, R.; Dubourg, V. Scikit-Learn: Machine learning in Python. *J. Mach. Learn. Res.* **2011**, *12*, 2825-2830.
(14) Pukrittayakamee, A.; Malshe, M.; Hagan, M.; Raff, L. M.; Narulkar, R.; Bukkapatnum, S.; Komanduri, R. Simultaneous Fitting of a Potential-Energy Surface and Its Corresponding Force Fields Using Feedforward Neural Networks. *J. Chem. Phys.* **2009**, *130*, 134101.
(15) Nguyen-Truong, H. T.; Le, H. M. An Implementation of the Levenberg–Marquardt Algorithm for Simultaneous-Energy-Gradient Fitting Using Two-Layer Feed-Forward Neural Networks. *Chem. Phys. Lett.* **2015**, *629*, 40-45.
(16) Gastegger, M.; Behler, J.; Marquetand, P. Machine Learning Molecular Dynamics for the Simulation of Infrared Spectra. *Chem. Sci.* **2017**, *8*, 6924-6935.
(17) Behler, J. Constructing High-Dimensional Neural Network Potentials: A Tutorial Review. *Int. J. Quantum Chem.* **2015**, *115*, 1032-1050.
(18) Bishop, C. *Pattern Recognition and Machine Learning*; Springer-Verlag: New York; 2006.
(19) Hu, D. P.; Liu, Y. F.; Sobolewski, A. L.; Lan, Z. G. Nonadiabatic Dynamics Simulation of Keto Isocytosine: A Comparison of Dynamical Performance of Different Electronic-Structure Methods. *Phys. Chem. Chem. Phys.* **2017**, *19*, 19168-19177.